\def \bfr {\begin{flushright}}
\def \efr {\end{flushright}}
\def \caja {\makebox[3.2cm][1]}
\def \v {\vskip}
\def \R {{\it R}}
\def \d {\hbox{d}\,}
\def \square {\hbox{$\sqcup\!\!\!\!\sqcap$}}

\def \e {\hbox{e}}

\def \p {\partial}

\def \w {\omega}

\def \ba {\begin{array}}
\def \ea {\end{array}}

\def \bea {\begin{eqnarray}}
\def \eea {\end{eqnarray}}

\def \be {\begin{equation}}
\def \ee {\end{equation}}

\def \bfr {\begin{flushright}}
\def \efr {\end{flushright}}
\def \caja {\makebox[3.2cm][1]}
\def \J {{\cal J}}
\documentstyle[12pt,a4]{article}
\begin{document}

%%%%%%%%%%%%%%%%%%%%%%%%%%%%%%%%%%%%%%%%%%%%%%%%%%%%%%%%%%%%%%%%%%
%%%%%%%%%%%%%%%%%%%%%PRINCIPIO DEL DOCUMENTO%%%%%%%%%%%%%%%%%%%%%%
%%%%%%%%%%%%%%%%%%%%%%%%%%%%%%%%%%%%%%%%%%%%%%%%%%%%%%%%%%%%%%
\thispagestyle{empty}
%\noindent{\bf Preliminary Version}
\noindent hep-th/yymmxxx\hfil\caja{{\bf Imperial-TP/94-95/8}}\break
%\today
\v20mm
\begin{center}

{\bf GROUP QUANTIZATION ON CONFIGURATION SPACE}
\footnote[1]{Work partially supported by the DGICYT.}

{\it Miguel Navarro\footnote[2]{On leave of absence
from [2]\ and [3].}$^{1}$,  V\'\i ctor Aldaya$^{2,3}$ and
Manuel Calixto$^{2}$}
\v5mm
\end{center}
\begin{enumerate}

\item The Blackett Laboratory, Imperial College, Prince
Consort Road, London SW7 2BZ; United Kingdom.

\item Instituto Carlos I de F\'\i sica Te\'orica y Computacional,
Facultad  de  Ciencias, Universidad de Granada, Campus de Fuentenueva,
18002, Granada, Spain.

\item IFIC, Centro Mixto Universidad de
Valencia-CSIC, Burjassot 46100-Valencia, Spain.
\end{enumerate}
\v5mm
\centerline{\bf Abstract}
\footnotesize
New features of a previously introduced Group Approach to Quantization
are presented.
We show that the construction of  the symmetry group
associated with the system to be quantized
(the ``quantizing group")
does not require, in
general, the explicit construction of
the phase space of the system, i.e., does not require the actual
knowledgement of the general solution of the classical
equations of motion: in many relevant cases
an implicit construction
of the group can be given, directly, on configuration space.
As an application we construct the symmetry group for
the conformally invariant massless scalar and electromagnetic fields and
the scalar and Dirac fields evolving in a symmetric curved space-
time or interacting with symmetric classical
 electromagnetic fields. Further generalizations of the present
procedure are also discussed and in particular the conditions under which
non-abelian (mainly Kac-Moody) groups can be included.
\normalsize
\vfil\eject
\setcounter{page}{1}

\section{Introduction.}

The convencional perturbative methods of quantization does not work
properly with several relevant field theories.
In addition to this, even in the case of theories for which a
perturbative approach is possible, there is some information which
cannot be obtained by perturbative techniques because of its global
nature. It is ,
therefore, necessary to look for other non-perturbative
methods to extract this information from these quantum theories.
A quantization method which might be specially suitable to perform
this task is the Group Approach to Quantization (GAQ) formalism.

The GAQ formalism was introduced several years
ago (see, e.g. \cite{[JMP],[CMP]} and references therein) as an improved
version of the
Geometric Quantization and the Kirillov
coadjoint orbit methods of quantization \cite{[Woodhouse],[Kirillov]}.
One of the major aims in the construction of the algorithm was the possibility
of arriving at the quantum solutions of a given physical system without
explicitly solving the corresponding classical equation of motion,
thus allowing for a quantum system for which a classical limit is not
properly defined or the classical equations do not have a well defined
general solution. However, the required understanding of the basic
symmetry group \footnote{In this paper we shall refer to this
group as a quantum group. The reader should not confuse this
notion  with another notion of ``quantum group"
 which frequently appears in the literature and which, in fact, does not
correspond to a group nor a quantum system either.}
is  so accurate that, in many cases, the effort to findig
it could be nearly equivalent to solving the problem.
In this sense, GAQ is perhaps more
useful as a tool for associating exactly
solved (quantum systems) with
already classified Lie groups than a method to quantize physical
systems originally defined by a Lagrangian.

This paper is intended to be a first step in the opposite direction,
i.e. towards the construction of quantum groups,
closely related to actual Lagrangians
(or, equivalently, equations of motions),
and depending on fields
given in configuration space, that is to say,
 formal groups of sections of a given
fiber bundle on space-time directly attached to the physical system.
We shall call this form of the formalism
{\it configuration-space image} of the formalism.
As basic result, previous to this goal, it will be shown that
the construction of the quantum
group does not require, in principle, the previous step of going
 to the phase space of the system, i.e., does
not require to
solve explicitly the classical equations of motion.

The configuration-image of the formalism will provide us in addition
a clearer view of the exact nature of the
quantum group and will make it clear the relationship
of this formalism with the Lagrangian
and canonical formalisms.

We shall also show some of
the advantages of expressing
the formalism in configuration-space image by constructing the quantum
group for several non-trivial fields in a
natural and straighforward way:
the conformally invariant massless escalar and
electromagnetic fields, and the scalar field in a symmetric
curved space-time or interacting with arbitrary symmetric
classical electromagnetic fields (Section 3).

This image of the formalism, as well as the examples here presented, can
also serve as a guide
for further generalizations. We will discuss on some of
them and will argue that a direct
generalization for non-abelian Kac-Moody groups requires the equations
of motion to be of first order.
We shall show by means of two examples
that this is not the case for other types of non-abelian
groups (Section 4).

In this paper we shall not get involved with the
subtleties of the proper quantization procedure. The interested
reader may found them in some of the quoted references.

\section{The harmonic oscillator and the Klein-\break  Gordon field.}
\label{a}

In this section we shall make use of two simples examples,
the harmonic oscillator
and the Klein-Gordon field, to present the basic features of the
formalism when the quantum group can be writen in terms of
fields in configuration espace.

The group law proposed in ref. \cite{[JMP]} for the one-dimensional
harmonic oscillator [the extended one-dimensional Galilei group,
the quantum group for the non-relativistic free particle can be
obtained as the limiting case $\w\rightarrow 0$] was:

\bea \label{a1}
A''&=&A + A'\cos \w B +\left(\frac{V'}\w\right)\sin\w B\nonumber\\
V''&=& V + V'\cos\w B - \w A'\sin \w B\\
B''&=& B + B'\nonumber\\
\zeta''&=&\zeta\zeta'\exp\frac{i}2\left[V(A'\cos\w B +\frac{V'}\w\sin\w B)-
A(V'\cos\w B - \w A'\sin \w B)\right]\nonumber\eea
Nevertheless, appart from the fact that the algebra of this group is
isomorphic to the algebra of the basic observables of the harmonic
oscillator not much has been explained on the manner
in which this group is actually associated with the harmonic oscillator.
Let us consider, however,
the H.O. from the lagrangian point of view. The action is:

\be\label{a2}
 S_{H.O.}=\frac12m\int\d t\left[\dot q^2-\w^2 q^2\right]\ee
The equations of motion are,  therefore,

\be\label{a3}
\left[\frac{\d^2}{\d t^2}+\w^2\right] q=0\ee
with general solution:
\bea\label{a4}
q(t)&=&q_0\cos\w t+\frac{\dot q_0}\w \sin\w t\nonumber\\
\Rightarrow\dot q(t)&=&\dot q_0\cos\w t-\w q_0\sin\w t\eea
Hence, by means of the identification $A\equiv
q_0\;,\;V\equiv\dot q_0\;,\;t\equiv B$; and associating
 a time evolution
 to the coordinates of the phase space in the natural manner, the group law
in eq. (\ref{a1}) can be written in the form:

\bea\label{a5}
A''&=& A+A'(B)\nonumber\\
V''&=& V+V'(B)\\
B''&=&B+B'\nonumber\\
\zeta''&=&\zeta\zeta'\exp\frac{i}2m\left[VA'(B)-AV'(B)\right]\nonumber\eea
Now, taking into account that $\dot q\equiv\dot A=V$,
this group law can straightforwardly
be written on configuration space:

\bea\label{a6}
B''&=&B+B'\nonumber\\
q''(t)&=&q(t)+q'(t+B)\\
\zeta''&=&\zeta'\zeta\exp\frac{i}2m\left[\dot q(t)q'(t+B)-
q(t)\dot q'(t+B)\right]_{|_{t=t_0}}\nonumber\eea

Instead of discussing here what can we learn from the simple
manipulations above and the results obtained from them,
it is preferable to consider first the case of a field,
the  Klein-Gordon field for instance. The action and equations of
motion are:

\bea
S_{KG}&
=&\frac12\int\d^4x\left[\p^\mu\phi\p_\mu\phi-m^2\phi^2\right]\label{b1}\\
\Rightarrow&&\left[\square+m^2\right]\phi=0\label{2b1}\eea

The analogy with the harmonic oscillator can serve us as a guide to
propose a quantum group for the fields directly on configuration space
(in order to simplify the discussion we
will not consider here Lorentz transformations,
but only the symmetries associated with
 space-time translations). Let us
try with the following composition law:

\bea\label{b4}
a''&=&a+a'\qquad\qquad a\in\R^4\nonumber\\
\phi''(x)&=&\phi(x)+\phi'(x+a)\\
\zeta''&=&\zeta'\zeta\exp\frac{i}2\xi_{KG}(g',g)\nonumber\\
&\equiv&\zeta'\zeta\exp\frac{i}2\int_{\Sigma}\d \sigma_\mu(x)
{\cal J}_{KG}^\mu\left(g,g'\right)(x)\nonumber\eea
where
\be\label{b5}
{\cal J}_{KG}^\mu\left(g,g'\right)(x)=
\p^\mu\phi(x)\phi'(x+a)-\phi(x)\p^\mu\phi'(x+a)
\ee
and $\Sigma$ is any space-like hypersurface.

Now, let us
 expand the fields $\phi$ that are solution of the
equation of motion (\ref{2b1}) in Fourier modes by means of

\bea \phi(x)=\int\frac{\d^3k}{2k^0}
\left(\Phi(k)\e^{-ikx}+\Phi^+(k)\e^{ikx}\right)\label{Fourier}\eea
It is not difficult to see that we get for the Fourier modes $\Phi(k)$
 the composition law postulated
in ref. \cite{[JPA]}.
This implies that the composition law above actually corresponds to
a group law. It is, nonetheless, easy
to see that we do not have to make use of the general solution
(\ref{Fourier}) to show that eqs. (\ref{b4},\ref{b5}) defines
 a group law.
In fact, the requirement that the fields in eq. (\ref{b5}) be solutions
of the equation of motion is enough to show that

 a) the quantity $\xi(g',g)$ fulfils the cocycle
property:

\bea \xi(g'',g')+\xi(g''*g',g)=\xi(g'',g'*g)+\xi(g',g)\label{cocicle}\eea
and that

b) the current ${\cal J}^\mu$ is conserved

\be\label{b6}
\p_\mu{\cal J}^\mu=0\ee

In fact, in all the cases studied in this paper
 there is a double implication  a) $\Leftrightarrow$ b):
to find out a divergenceless current made up of the fields solution of
the equations of motion has always proved to be sufficient for the quantity
$\xi$, constructed on it,
to fulfil the cocycle property.
We do not know whether this is the general case.

We can, therefore, state the following features of the formalism when
 expressed in configuration space:

1. The basic group (the group to be centrally extended) is a group
irrespective whether or not the fields involved
fulfil the equations of motions.

2. The central extension involves the integral of a divergenceless
current ${\cal J}^\mu$ over an hypersurface $\Sigma$. This current is
divergenceless only over the fields that obey the equations of
motion.

Therefore the centrally extended group, the quantum group,
{\it involves only the fields that are solutions of the equations of
motion}.  It is constructed upon the {\it phase space} of the system.

In this phase space different coordinates can, of course, be chossen.
The choice
of the Fourier modes $\Phi(k)$ leads us to the composition law in ref.
\cite{[JPA]}. Another choice is the familiar one of
fields and time-like derivatives of the fields
in the hypersurface $\Sigma$. We can write down the group law in
these coordinates by making use of the following
propagation property:

\be\label{b2}
\phi(y)=\int_{\Sigma}\d\sigma_\mu(x)\left[\p^\mu\Delta(y-x)\phi(x)
-\Delta(y-x)\p^\mu\phi(x)\right]\ee
where the propagator $\Delta$ obeys the equation of motion
\cite{[Ryder]}:

\bea\label{b3}
\left[\square+m^2\right]\Delta=0\eea

3. The classical equations of motion, {\it but not their general
solution}, are required to show that  the
quantum group is, in fact, a group.

4. The equation of motion are not determined uniquely by the
group: eq. (\ref{b6}) implies eq. (\ref{2b1}) for {\it some}
$m$ but not for a particular $m$.
Therefore
eq. (\ref{b6}) {\it almost} implies eq. (\ref{2b1}) but there is not a
complete implication.
 The groups in configuration space
for the harmonic oscillator and the free particle are the same
whereas the equations of motion are not. It is in terms
of phase-space coordinates when the difference between these systems
explicitly appears in the group.

\section{Applications.}

In this section we will construct the quantum group for several physically
interesting systems
as an application of the formalism presented above:
the conformally-invariant massless scalar
and electromagnetic fields, the scalar and Dirac field in a symmetric
curved space-time and these same fields interacting with symmetric
electromagnetic fields.

[Unlike in the
previous section, where the semidirect action of the space-time
symmetries
were given in a rather unnatural way - a way adapted
to the Schr\"odinger representation -
we shall write in the sequel the quantum groups in the natural
way; the other expression can be obtained by a simple change of
coordinates.]

\subsection{Conformally invariant fields.}

        The conformal group, whose composition law have not, up to now,
been given in a closed form, is made up of compositions of
the following actuations on the space-time:

$\ba{ll} \hbox{a) Space-time translations:}& (ux)^\alpha = x^\alpha
+ a^\alpha\\
\hbox{b) Lorentz transformations:}&(ux)^\alpha=\Lambda^\alpha_\mu
x^\mu\\
\hbox{c) Dilatations:}& (ux)^\alpha=\e^\lambda x^\alpha\\
\hbox{d) Special conformal transformations:}&
         (ux)^\alpha=\frac{x^\alpha +c^\alpha x^2}
        {1+2cx+c^2x^2}\ea$

\subsubsection{The massless scalar field.}
        The group law (in four dimensions, in which case the conformal
dimension
of a scalar field is $l=-1$) is:

        \bea u''&=&u'*u\qquad\hbox{Conformal (sub)group}\nonumber\\
        \phi''(x)&=&\phi'(x)+\Omega^{-1}(u'^{-1},x)\phi(u'^{-1}(x))
        \label{n2}\\
\zeta''&=&\zeta'\zeta\exp\frac{i}2\xi_{MS}(g',g)\nonumber\\
       &=&\zeta\zeta'\hbox{exp}\>{{\hbox{i}}\over2}\int_\Sigma
        d\sigma_\mu(x) {\cal J}_{MS}^\mu\label{n3}\eea
\noindent where
        \bea
        {\cal J}_{MS}^\mu=\phi'(x)\partial^\mu
        [\Omega^{-1}(u'^{-1},x)\phi(u'^{-1}(x))]-
        \partial^\mu\phi'(x)
        [\Omega^{-1}(u'^{-1},x)\phi(u'^{-1}(x))]\label{n4}\eea
\noindent  and the function $\Omega$ is given by:
\v 3mm
$\Omega(u,x) = \left\{
\ba{cl} 1+2cx+c^2x^2
&\hbox{for special conformal transformation,}\\
\e^{-\lambda}&\hbox{for dilatations,}\\
1&\hbox{for the Poincar\'e subgroup.}\ea\right.$
\v 3mm

\noindent The function $\Omega$ for a general conformal transformation can
be obtained by making use of its property

\bea \Omega(u,x)\Omega(u',ux)=\Omega(u'u,x)\label{n5}\eea
which is required for eq. (\ref{n2}) to define a group.

     It is not difficult to show that if
$\phi(x)$ and $\phi'(x)$ are solutions of the equation of motion

\be \partial^\mu\partial_\mu\phi(x)=0\label{n6}\ee
so is $\phi''(x)$.
It is also straighforward to show, by making use of eqs. (\ref{n5}-
\ref{n6}), that the current ${\cal J}^\mu$ is divergenceless and
fulfils the cocycle property (\ref{cocicle}).

        \subsubsection{The electromagnetic field.}
The group law is given by:

        \bea\label{m1}
        u''&=&u'*u\qquad\hbox{Conformal (sub)group}\nonumber\\
        A''_\mu(x)&=&A'_\mu(x)+
        \frac{\p u'^{-1\alpha}}{\p x^\mu}A_\alpha(u'^{-1}x)\\
        &\equiv&A'_\mu(x)+(S(u'^{-1})A)_\mu(x)\nonumber\eea
where $S$ is the representation of the conformal group that acts
on the electromagnetic  vector field. This actuation is the
natural one and means that the potential vector has null
conformal weight.  This actuation induces the following one on
the tensor field $F_{\mu\nu}$:

        \bea\label{m2}
        F''_{\mu\nu}(x)&=&F'_{\mu\nu}(x)+\frac{\p u'^{-1\alpha}}{\p x^\mu}
        \frac{\p u'^{-1\beta}}{\p x^\nu}F_{\alpha\beta}(u'^{-1}x)\\
        &\equiv&F'_{\mu\nu}(x)+(S(u'^{-1})F)_{\mu\nu}(x)
        \eea
It is easy to show that this actuation leaves invariant
Maxwell's action

        \bea S_M=\int\>\d^4x\> F_{\mu\nu}\>F^{\mu\nu}\label{m3}\eea
and, therefore, leaves invariant  Maxwell's equations.

The central extension is given by:

        \bea\label{m4}
\zeta''&=&\zeta'\zeta\exp\frac{i}2\xi_M(g',g)\nonumber\\
        &=&\zeta'\zeta\exp\frac{i}2
        \int_{\Sigma}\d \sigma_\mu(x)
        {\cal J}^\mu_M\left(g',g\right)(x)\eea

\noindent with divergenceless current

\bea\label{m5}
        {\cal J}_M^\mu\left(g',g\right)(x)
     =F'^{\mu\nu}(x)(S(u'^{-1})A)_\nu(x)-
        A'_\nu(x)(S(u'^{-1})F)^{\mu\nu}(x)\eea

        If we restrict ourselves to the symmetry group of space-time
translations the group is written:

\bea\label{c1}
a''&=&a+a'\nonumber\\
A_\mu''(x)&=&A'_\mu(x)+A_\mu(x-a')\\
\zeta''&=&\zeta\zeta'\exp\frac{i}2
\int_{\Sigma}\d \sigma_\mu(x)
{\cal J}_M^\mu\left(g',g\right)(x)\nonumber\eea
with a current

\be\label{c2}
{\cal J}_M^\mu\left(g',g\right)(x)=F'^{\mu\nu}(x)A_\nu(x-a')-
A'_\nu(x)F^{\mu\nu}(x-a')\ee

Here we can see again, in this last example,
the issue already notized above: The
current $\J_M^\mu$ is divergenceless and fulfils the cocycle
property if Maxwell equations are obeyed, but a Proca-like
equation $\p_\mu F^{\mu\nu}+m^2 A^\nu=0$, with a non-null mass
$m$, is also allowed. This is, however, no longer the case when
the full conformal group is considered.

\subsection{Matter fields in a symmetric curved space-time.}
In this section we will present the quantum group for matter fields,
the Klein-Gordon and  Dirac fields, evolving in a symmetric, but on the
other hand arbitrary, curved space-time with metric $g_{\mu\nu}$.

Let us suppose that the set of transformations $v$ is a group
of isometries of the metric (see for instance
refs. \cite{[Nakahara],[Marias]}). Then we have

\bea g_{\mu\nu}(vx)\frac{\p (vx)^\mu}{\p x^\alpha}\frac{\p (vx)^\nu}
        {\p x^\beta}=g_{\alpha\beta}(x)\label{killing1}\eea
and, in the same way,

        \bea g^{\mu\nu}(vx)=g^{\alpha\beta}(x)\frac{\p (vx)^\mu}{\p
        x^\alpha}\frac{\p (vx)^\nu}{\p x^\beta}\label{killing2}\eea

\subsubsection{The scalar field.}

The equations of motion for a scalar field evolving in this
background metric are (see for instance ref. \cite{[Birrell],[Fulling]}):

\bea \left[\square(x) +\alpha R(x)+m^2 \right]\phi(x)=0\label{sce1}\eea
with

\bea \square(x)\phi(x)\equiv
        \frac1{\sqrt{g(x)}}\p_\mu\left(\sqrt{g(x)}g^{\mu\nu}(x)
        \p_\nu\right)\phi(x)\eea

The group law that would describe the quantum dynamics of this
system is given by:

        \bea v''&=&v'*v\nonumber\\
        \phi''(x)&=&\phi'(x)+\phi(v'^{-1}(x))
        \label{killing4}\\
        \zeta''&=&\zeta\zeta'\hbox{exp}\>{{\hbox{i}}\over2}\int_\Sigma
        d\sigma_\mu(x) {\cal J}_{SCS}^\mu\label{killing5}\eea
\noindent with
        \bea
        {\cal J}_{SCS}^\mu=\sqrt{g(x)}g^{\mu\nu}(x)
        \left[\phi'(x)\partial_\nu\left[\phi(v'^{-1}(x))\right]-
        \partial_\nu\phi'(x)\phi(v'^{-1}(x))\right]\label{killing6}\eea
(Notize that eq. (\ref{killing1}) implies
$\square(x)=\square(vx)$ and $R(vx)=R(x)$
(see also ref. \cite{[Marias]}).)

\subsubsection{The Dirac field.}

Let now $\Psi$ be a Dirac field with equations of motion
(see e.g. ref. \cite{[Nakahara],[Birrell]}):

\be \left[\hbox{i}\hat\gamma^\mu\left(\p_\mu +
\frac12\hbox{i}{{\Gamma^a}_\mu}^b\Sigma_{ab}\right)
-m\right]\Psi=0\label{dce1}\ee
where

\bea g_{\mu\nu}&=& \eta_{ab}\,\e^a_\mu\e^b_\nu\label{dce2}\\
 \hat\gamma^\mu &=&\gamma^a{\e_a}^\mu\quad,\quad
\Sigma_{ab}=\frac14\hbox{i}\left[\gamma_a\,,\,\gamma_b\right]\\
\Gamma^c_{ab}&=&{\e^c}_\nu\e_a^\mu
\left(\p_\mu{\e_b}^\nu + {\e_b}^\lambda\Gamma^\nu_
{\mu\lambda}\right)\eea

The transformations $v$ being isometries implies that there
exist a set of local Lorentz transformations $\Lambda(v,x)$ such that:

\bea {\e^a}_\mu(vx)\frac{\p (vx)^\mu}{\p x^\lambda}
={\Lambda(v,x)^a}_c\,{\e^c}_\lambda(x)\label{dce14}\eea

On this grounds it can be shown that
the following set of transformations is a (super)group:

 \bea v''&=&v'*v\nonumber\\
        \Psi''(x)&=&\Psi'(x)+\rho(\Lambda(v'^{-1},x))\Psi(v'^{-1}(x))
        \label{dce22}\\
\bar\Psi''(x)&
=&\bar\Psi'(x)+\bar\Psi(v'^{-1}(x))\rho(\Lambda(v'^{-1},x))^{-1}\nonumber\\
\zeta''&=&\zeta'\zeta\exp\frac{i}2\xi_{DCS}(g',g)\nonumber\\
        &=&\zeta'\zeta\exp\frac{i}2
        \int_{\Sigma}\d \sigma_\mu(x)
        {\cal J}_{DCS}^\mu\left(g',g\right)(x)\label{dce23}\eea

\noindent with
        \bea\label{dce24}
       {\cal J}_{DCS}^\mu\left(g',g\right)(x)
&=&\hbox{i}\left[\bar\Psi'(x)\hat\gamma^\mu(x)\rho(\Lambda(v'^{-1},x))\Psi(v'^{-1}x)\right.\nonumber\\
&&\left.-
\bar\Psi(v'^{-1}x)\rho(\Lambda(v'^{-1},x))^{-1}
\hat\gamma^\mu(x)\Psi'(x)\right]\eea
and $\rho$ the usual spin representation of the Poincar\'e group which
verifies:

\be \rho(\Lambda)^{-1}\gamma^a\rho(\Lambda)={\Lambda^a}_b\gamma^b
\label{dce26}\ee

\subsection{Matter fields coupled to symmetric electromagnetic fields.}
In this subsection we will present the quantum group for matter fields,
the Klein-Gordon and  Dirac fields, coupled to a symmetric, but on the
other hand arbitrary, electromagnetic field.

The espace-time in this
section will be flat with a Minkowskian metric $\eta_{\mu\nu}$.
The set of transformation $v$ will be any subgroup of the Poincare
group leaving invariant
the electromagnetic field: $A_\mu(vx)=A_\mu(x)$.

\subsubsection{The Klein-Gordon field.}

Let the scalar field $\phi$ obey the equation of motion

\be \left(D_\mu\,D^\mu +m^2\right)\phi=0\label{af1}\ee
with $D_\mu\equiv \p_\mu-iA_\mu$.

Then the
following composition law defines a group:

    \bea v''&=&v'*v\label{af3}\\
        \phi''(x)&=&\phi'(x)+\phi(v'^{-1}(x))
        \label{af4}\\
\zeta''&=&\zeta'\zeta\exp\frac{i}2\xi_{SAF}(g',g)\nonumber\\
        &=&\zeta'\zeta\exp\frac{i}2
        \int_{\Sigma}\d \sigma_\mu(x)
        {\cal J}_{SAF}^\mu\left(g',g\right)(x)\label{af5}\eea
\noindent with

\bea {\cal J}_{SAF}^\mu\left(g',g\right)(x)
&=&D^\mu\phi(v'^{-1}){\phi^*}'(x)-\phi(v'^{-1}){(D^\mu\phi)^*}'(x)\
\nonumber\\
&&+\left(D^\mu\phi(v'^{-1})\right)^*\phi'(x)-
\phi(v'^{-1})^*(D^\mu\phi)'(x)\label{af7}\eea

\subsubsection{The Dirac field.}

Let now $\Psi$ be a Dirac field with equations of motion:

\be \left(\hbox{i}\gamma^\mu D_\mu-m\right)\Psi=0\label{af8}\ee
Then the following set of transformations is a group:

 \bea
v''&=&v'*v\nonumber\\
\Psi''(x)&=&\Psi'(x)+\rho(\Lambda(v'^{-1}))\Psi(v'^{-1}(x))\label{af10}\\
\bar\Psi''(x)&=&\bar\Psi'(x)+
\bar\Psi(v'^{-1}(x))\rho(\Lambda(v'^{-1}))^{-1}\nonumber\\
\zeta''&=&\zeta'\zeta\exp\frac{i}2\xi_{DAF}(g',g)\nonumber\\
        &=&\zeta'\zeta\exp\frac{i}2
        \int_{\Sigma}\d \sigma_\mu(x)
        {\cal J}_{DAF}^\mu\left(g',g\right)(x)\label{af13}\eea
\noindent with

        \bea\label{af20}
       {\cal J}_{DAF}^\mu\left(g',g\right)(x)
&=&\hbox{i}\left[\bar\Psi'(x)\gamma^\mu\rho(\Lambda(v'^{-1}))\Psi(v'^{-1}x)\right.\nonumber\\
&&\left.- \bar\Psi(v'^{-1}x)\rho(\Lambda(v'^{-1}))^{-
1}\gamma^\mu\Psi'(x)\right]\eea

\subsection{Comments.}
We should point out here that
the requirement of being symmetric for the
classical fields considered above,
such as the space-time metric or the
electromagnetic field in subsection 3.2 and 3.3 respectively,
is not very restrictive a requirement:
many interesting systems, such
as the one of fields evolving
in a Schwarzchild black-hole background \cite{[Birrell],[Wald]},
are not excluded by these constraint.
An analysis in depth of some of these systems is in progress
and will appear elsewhere.

\section{Generalizations. The Virasoro group and the Schwarzian derivative.}
\label{v}

        Let us consider from another point of view
the manner in which  we  arrived  at
the group law for the harmonic oscillator in Sec. 2, eq. (\ref{a1}) [Similar
consideration can be made on the group laws for the other fields
mentioned above.] We  started with an abelian Kac-Moody group (the
group of loops on $\R$) [composed in a semidirect way with the
temporal translations which are the zero modes of the Virasoro group on
the real line]. From this group we  extracted the subgroup of the
functions that obey certain differential equations (the equations of
motion). The subgroup obtained this way can be extended, the
principal ingredient for the extension being a divergenceless
current.

This point of view leads us inmediately to a possible generalization:
in place of an abelian Kac-Moody group let us consider more general,
non-abelian Kac-Moody groups.
But, what are the differential equations (equations of motion)
to be applied to the basic
group? In another words: what differential equation is there
such that if $g(t)$ and $h(t)$ are solutions of this equation
also is $g(t)*h(t)$ for a general non-abelian Kac-Moody group?.
(This is a sort of symmetry that goes beyond those usually considered
in the literature.)
 For the abelian Kac-Moody group the solution is
obvious: any linear diferential equation is such that if $h(t)$,
$g(t)$ are solutions so is $h(t)+g(t)$. From this simple property
we can construct the quantum group for the harmonic oscillator,
the non-relativistic free-particle,
the Klein-Gordon and Maxwell fields...
But, what differential equation fulfils this property for, say, a
$SU(2)$-Kac-Moody group?

There is strong indications that no equation of order greater than one
with this property exists for any
 non-abelian Kac-Moody group. In fact, let us
consider the Lie algebra of these Kac-Moody groups.
For any abelian Kac-Moody
group, the basic Lie algebra is:

\bea \left\{\varphi_a(x),\varphi_b(y)\right\}=0\label{v1}\eea
{}From this algebra we have to extract the
subalgebra made up of the tangent fields that obey
the (linearized) equations of motion and further extend it.
In the abelian case this can be done with the result

\bea \left\{\varphi_a(x),\varphi_b(y)\right\}=\Delta_{ab}(x-y)\label{v2}\eea
where $\Delta_{ab}$ is the propagator similar to that of eqs.
(\ref{b2}-\ref{b3}).
[If $\Delta(x-y)$ is the propagator for the Klein-Gordon field, the
propagator for the Dirac field is
$\Delta_{D}(x-y)=\left(\hbox{i}\gamma^\mu\p_\mu+m\right)\Delta(x-y)$ and for
the electromagnetic (or Proca) field is
${\Delta_{M}}_{\mu\nu}(x-y)=  -\eta_{\mu\nu}\Delta(x-y)$.
For the harmonic oscillator it is $\Delta_{HO}(B)=\frac1\w\sin\w B$ and
for the non-relativistic free particle  $\Delta_{FP}(B)=B$.]
This central extension is consistent with the (linearized) equations
of motion for the tangent fields.

For non-abelian fields we should start from the basic Lie algebra:

\bea \left\{T^a(x),T^b(y)\right\}=f^{ab}_cT^c(x)\delta(x-y)\label{v3}\eea

The Lie bracket of two elements of the Lie algebra

\be X=\int \d\,x\;f_a(x)T^a(x)\; ,\;Y=\int\d\,x\;g_a(x)T^a(x)\ee
is given by

\be \left[X\;,\;Y\right]=\int
\d\,x\;f_a(x)g_b(x)f^{ab}_c\;T^c(x)\label{v4}\ee
Therefore the Lie algebra (\ref{v3}) can equivalently be written
in terms of the coefficient functions $f$ as follows:

\be  \left[f\;,\;g\right]_c(x)
=f_a(x)\,g_b(x)f^{ab}_c\label{v5}\ee

For any equation of motion that we impose now on the group
elements, the induced linearized equations of motion for the
elements of the Lie algebra will be, of course, linear.
But, for non-null structure constants $f^{ab}_c$, eq. (\ref{v5})
implies that no subalgebra can be defined by linear equations of order
greater than one.

This result is a sort of {\it no go} theorem
which implies that the construction above
for abelian bosonic groups (fields), with equations of motion of
second order, cannot be {\it directly} extended
to non-abelian Kac-Moody groups. This extension would require
expressing the equations of motion in a typically Hamniltonian form.
Only in the case of abelian Kac-Moody groups the splitting of
coordinate-momentum can be made whithout breaking the group structure.

Now, we will show that this obstruction is not present in other kind of
non-abelian groups.  Let us consider, as an example,
the Schwarzian derivative $S(f)$:

\bea S(f)=\p_t\left\{\frac{\p^2_tf}{\p_tf}\right\}
-\frac12\left(\frac{\p^2_tf}{\p_tf}\right)^2=\frac{\p^3_tf}{\p_tf}
-\frac32\left(\frac{\p^2_tf}{\p_tf}\right)^2\label{s1}
\eea
This operator fulfils the so called Cayley property:

\bea
S(fog)=S(f)og(\dot g)^2+S(g)\label{s2}\eea
where $o$ stands for composition of functions.
Therefore we will have that

\bea S(f)=0\;\;,\;\; S(g)=0\quad\Longrightarrow\quad S(fog)=0.
\label{s3}\eea
Therefore, for the Virasoro group (the group of loops defined in a one
dimensional manifold with the composition of function as its
composition law),  the vanishing of the Schwarzian derivative
is a differential equation
with the property we were looking for. (We will say that the
Schwarzian derivative is {\it closed} under the Virasoro group.)
Since the Virasoro group {\it is not} a Kac-Moody group,
it does not fulfil the hipotesis of the no-go theorem above, and
we see that it does fulfil its tesis neither.

The algebra of the loop group on $\R$ is the Lie algebra of all
vector fields $v(t)\p_t$  with Lie bracket:

\bea \left\{u,v\right\}(t)=-(\dot u(t)v(t)-u(t)\dot v(t))\label{s4}\eea
If the functions of the group satisfy $S(f)=0$, by taking variation
and taking into account that the identity function is $f(t)=t$,
we arrive at the following (of course linear) equation for the
functions in the Lie algebra of the group

\bea \p^3_tv(t)=0\;.\label{s5} \eea
It is straighforward to show b applying the lineal operator
$\p^3_t$ to both sides of eq. (\ref{s4}) that the set of functions which
satisfy eq. (\ref{s5}) close into a subalgebra.

The general solution of equation (\ref{s5}) is a general linear
combination of $1,t,t^2$ which generates the general solution of
the equation $S(f)=0$:

\bea f(t)=\frac{at+b}{ct +d}\;,\quad\hbox{with} \quad ad-bc=1\label{s10}\eea
Therefore the subgroup of the Virasoro group generated  by
functions obeying $S(f)=0$ is the $SL(2,\R)$ group.

\subsection{The Virasoro-Kac-Moody Group and the induced 2D
gravity in the light-cone gauge.}
\v3mm
A construction similar to the one presented above for the
Virasoro group can also be made for the
{\it Virasoro-Kac-Moody group}.
[Do not get confused with the Kac-Moody-Virasoro groups which are
nothing other but the natural semidirect product
of the Virasoro group by a Kac-Moody group].
The  Virasoro-Kac-Moody group is the set of functions

\be\ba{rcl}f:\R&\longrightarrow &Virasoro\\
\sigma&\rightarrow&f(\sigma)\ea\label{vk1}\ee
with composition law $(f*g)(\sigma)(\tau)=(f(\sigma)og(\sigma))(\tau)$.
The elements of this group are, therefore, parametrized by two
coordinates $\sigma,\tau$ and can be considered as functions of two
variables, $f(\sigma,\tau)$.

The vanishing of the Schwarzian derivative
\be \left\{f(\sigma,\tau)\;,\tau\right\}=0\label{vk2}\ee
is,  of course,  a {\it closed} equation for this group, thus defining
a subgroup, the $SL(2,\R)$-Kac-Moody group in this case.

The identifications
$\sigma\equiv x^+=t+x\,,\,\tau\equiv x^-=t-x$ transform eq. (\ref{vk2})
into the equation of motion, $\left\{\,f\,,\,x^-\,\right\}=0$, of
the induced 2D gravity in the light-cone gauge \cite{[induced]}.
Moreover, in this gauge, the symplectic form is the canonical one
of the $SL(2,\R)$-Kac-Moody (sub)group. All this would indicate
that the quantum theory of this system should be described in terms of
irreducible, unitary
representations of the $SL(2,\R)$-Kac-Moody group.
However, it is well known that there are no unitary, standard
highest-weight representations with a nonzero central charge.
In fact, a more rigorous analysis of this theory based on a local
form of the
action shows that its true reduced phase space  is quite
different from that one and cannot be
obtained as a constrained version of the former \cite{[induced]}.
Therefore, the role played by the
$SL(2,\R)$-Kac-Moody symmetry is in any case different from the
one expected from the analysis of the non-local theory in the
light-cone gauge.

\section{Discussion.}

We have presented the basic features of the GAQ formalism when the
quantum group can be written in terms of fields given on
configuration space. The quantum group for several physically
meaningfull systems have been given, albeit most of them
in an implicit form.

But one might wonder what is the utility of writting down the quantum group in
this implicit form. The key point in considering the quantum group is that it
collects all the
information about both the classical and the quantum theories.
Of course we can do the most with it when  we know the general solution of the
equations of motion, i.e. the phase space of the theory
 (and, in fact, there are many important examples
in which this is the case) but we believe (and this issue
will be investigated in
forthcoming publications) that, even when the general solution is
not known, there can still be a lot of information
of the quantum theory that we can possibly extract from the quantum group.

An interesting case occurs when we only know {\it some} of the
classical solutions of the theory. With the GAQ
formalism we would be able to quantize these solutions provided that we were
able to find out pairs of solutions that are  coordinate-
momentum conjugate of each other.
(In the GAQ formalism, the cocycle may be
viewed as a sort of ``symplectic product"  which measures the extent
to which a pair of classical solutions are
coordinate-momentum conjugate of each other). This sort of {\it minisuperspace
approach}  would provide us with a first draft
of the quantum theory and
deserves a study of their own.

In addition to this, and in some sense, the phase space of any
classical theory is roughly known: it is
characterized by the fields and
time derivative of the fields in a Cauchy hypersurface and
evolving in accordance with the classical equations of motion.
Putting aside topological or, in general,
global issues of the phase space,
the GAQ formalim equiped with this rough description of
the phase space must necessarily work better than the familiar
Canonical Quantization do.

Furthermore, the configuration-space image of the GAQ formalism
appears as a well-suited formalism to deal with gauge theories.
Indeed, gauge symmetries, far from  being a mere useful tool to solve a
previously given theory, determine themselves the theory and this
philosophy is quite the same
that inspirates the GAQ formalism. However,
we  have proved in Sec. 4 a sort of no-go theorem which delimites the
type of equations of motion of a (non-linear)
system with a non-abelian Kac-Moody
symmetry: these equations must be kept in a Hamiltonian-like, first-order form,
since the restriction to the pure coordinate space satisfying second-order
equations would destroy the group structure. This situation resembles that of
WZW models \cite{[WZW]} where the Kac-Moody
symmetry comes out in a natural way when
written in a set of coordinates (light-cone coordinates) where the Lagrangian
is not regular, making this way the difference between
the Modified and Ordinary Hamilton Principle.
\v 1cm
\noindent{\bf Acknowledgements.} M. Navarro is grateful to the
Spanish MEC for a postdoctoral FPU grant.
M. Calixto thanks the Spanish MEC for a FPU grant.


\begin{thebibliography}{99}

\bibitem{[JMP]} V. Aldaya and J.A. de Azcarraga, {\it J. M. Phys.} {\bf
23}(7) (1982)1297-1305.

\bibitem{[JPA]} V. Aldaya, J.A. de Azcarraga and S. Garc\'\i a, {\it
J. Phys.} {\bf A21} (1988)4265-4287.

\bibitem{[CMP]} V. Aldaya, J. Navarro-Salas and A. Ram\'\i rez,
{\it Comm. Math. Phys.} {\bf 121},  (1989)541.

\bibitem{[Birrell]} N.D. Birrell and P.C.W. Davies, {\it Quantum
Fields in Curved Space}, Cambridge: Cambridge University Press,
(1982).

\bibitem{[Marias]} Y. Choquet-Bruhat, C. DeWitt-Morette and
M. Dillard-Bleick, {\it Analysis, Manifolds and Physics},
Asterdam-New York-Oxford: North-Holland Publishing Company (1977).

\bibitem{[Fulling]} S.A. Fulling, {\it Aspect of Quantum Fields
in Curved Space}, Cambridge: Cambridge University Press (1991).

\bibitem {[Kirillov]} A.A. Kirillov, {\it Elements of the Theory of
Representations} Springer, Berlin (1976).

\bibitem{[Nakahara]} M. Nakahara, {\it Geometry, Topology and
Physics}, Brixtol and New York: Adam Hilger (1991).

\bibitem{[induced]} J. Navarro-Salas, M. Navarro and V. Aldaya,
{\it Nucl. Phys.} {\bf B403} (1993)291-314

\bibitem{[Ryder]} L.H. Ryder, {\it Quantum Field Theory},
Cambridge: Cambridge University Press (1985).

\bibitem {[Woodhouse]} D. J. Simms and N. M. Woodhouse, {\it Lectures
in Geometric Quantization}, Springer-Verlag, Berlin, 1976; N. Woodhouse
{\it Geometric Quantization}, Clarendon, Oxford, (1980).

\bibitem{[Wald]} R.M. Wald, {\it General Relativity} Chicago and
London: The University of Chicago Press (1984).

\bibitem{[WZW]} E. Witten, {\it Comm. Math. Phys.} {\bf 92}, (1984)455;
 P. Goddard and
D. Olive, {\it Int. J. Mod. Phys.} {\bf A1}, (1986)303.

\end{thebibliography}
\end{document}